\documentclass[twocolumn,seceq]{jpsj2}

\def\runtitle{Two-Site shift Product Wave Function Renormalization Group Method
Applied to Quantum Systems}
\def\runauthor{H.~{\sc Ueda}, T.~{\sc Nishino}, and K.~{\sc Kusakabe}}

\title{Two-Site shift Product Wave Function Renormalization Group Method
Applied to Quantum Systems}

\author
{Hiroshi {\sc Ueda}$^1$, Tomotoshi {\sc Nishino}$^2$, and Koichi {\sc Kusakabe}$^1$  }

\inst
{
$^{1)}$Department of Material Engineering Science, 
Graduate School of Engineering Science, Osaka University, Osaka 560-8531\\
$^{2)}$ Department of Physics, Graduate School of Science, Kobe University, Kobe 657-8501
}

\recdate{\today}

\abst
{We report a way of wave function estimation for the density
matrix renormalization group (DMRG) method applied to quantum systems,
which has 2-site modulation, when the system size extension is necessary in
both the finite and the infinite algorithms.
The estimation is performed by renormalization group (RG) transformation 
applied to the ground state wave function, which is represented as the matrix product.
This RG scheme is known as the product wave function renormalization 
group (PWFRG) method. In order to treat the 2-site modulation, the operation 
of the RG transformation is shifted by amount of 2 lattice sites. It turns out that 
this 2-site shift algorithm provides better wave function estimation 
in the thermodynamic limit, compared with the previously known PWFRG algorithm.}

\kword{DMRG, Matrix Product Wave Function, Tensor Network, Renormalization Grup, }

\begin{document}
\sloppy
\maketitle

\section{Introduction}

Variational estimation of minimum eigenvalues of quantum Hamiltonians and 
maximum eigenvalues of classical transfer matrices has been investigated as a 
non perturbative way of analysis in condensed matter systems. The Kramers-Wannier
approximation applied to the two-dimensional (2D) Ising model is one of
the early example.~\cite{KW,Kikuchi} Baxter extended this formalism
by introducing auxiliary variable, and established the way of corner transfer 
matrix.~\cite{Baxter1,Baxter2,Baxter3} In the field of 1D quantum spin
system, the variational estimation of the ground state energy by Nightingale
and Bl\"ote is one of the earliest trial.~\cite{NB} Quantum
state constructed as a product of local factors occasionally represent exact 
ground state, or are good variational states.~\cite{AKLT,Klumper1,Klumper2}
Such states are known as {\it the matrix product state} (MPS), or {\it the finitely correlated
state}.~\cite{Fannes1,Fannes2,Fannes3} 
%
%
Practical and flexible use of the MPS for eigenvalue problems
began with the density matrix renormalization group (DMRG) 
method,~\cite{DMRG} which has been applied to
various problems in low dimensional correlated systems.~\cite{DMRG2,DMRG3}
The variational structure in DMRG formalism mediaged by MPS is 
revealed by \"Ostlund and Rommer.~\cite{Ostlund,Ostlund2,Ostlund3,Takasaki,Sierra}

It is known that numerical calculation in DMRG method can be accelerated by explicit
use of the matrix product structure of the variational state, 
especially when the method is applied to finite size
1D quantum systems.~\cite{Acce,Acce2} This acceleration procedure can be 
regarded as renormalization group (RG) transformation applied to 
the ground state wave function. It is possible to introduce 
this way of RG transformation to the infinite system 
DMRG method, where the acceleration procedure is named as `the product wave function 
renormalization group (PWFRG) method' since the RG transformation is
applied to the matrix product wave function.~\cite{PWFRG,PWFRG2}
Numerical efficiency of the PWFRG method is achieved by estimating a
trial wave function for the iterative calculation in the infinite system
DMRG method, and has been confirmed through 
applications to classical systems~\cite{Acts,Acts3} as well as quantum 
spin systems.~\cite{Hieida,Hagiwara,Okunishi,Narumi,Yoshikawa} 
It should be noted that the wave function estimation in the PWFRG method
is of use for the finite system DMRG method,~\cite{PWFRG2} when the 
system size increase is necessary for preparing numerical data for several
system sizes.

In this article we report an extension of the PWFRG method, which can be
applied to quantum systems that have 2 site modulation. According to this 
modulation, the RG transformation to the wave function is shifted by 2 lattice 
sites. This modified PWFRG method provides good wave function estimation 
when the infinite system DMRG method is nearly converges to the 
thermodynamic limit. We also discuss how to apply the PWFRG method
to matrix product wave functions obtained by the finite system DMRG method.

In the next section we explain the matrix product structure of the 
ground state wave function. In \S 3 we estimate the wave function, applying the RG
transformation to the ground state wave function. We check the numerical efficiency of the
estimated wave function in \S 4, where  fidelity error in the estimation process is observed. 
When there is finite excitation gap the wave function estimation works efficiently. 
We discuss the estimation scheme proposed by McClloch
quite recently, which provides better estimation than the PWFRG method especially
when the system is gapless.~\cite{McClloch}
Conclusions are summarized in the last section.

\section{Matrix Product Formulation}

Consider the eigenvalue problem for the ground state of a 1D quantum system
that has modulation of period 2. An example of such systems is the dimerized $S = 1/2$ 
Heisenberg spin chain of length $2N$, which is defined by the Hamiltonian
\begin{equation}
H^{( 2N )}_{~} = J \sum_{i = 1}^{2N - 1} \left\{ 1 + \delta ( - 1 )^i_{~} \right\} 
{\bf S}_i^{~} \cdot {\bf S}_{i+1}^{~} \, ,
\end{equation}
where $J > 0$ represents the antiferromagnetic interaction and 
where $\delta$ the dimerization. Since we treat
the MPS constructed by the infinite system DMRG method, the state which can be 
further improved by the finite system DMRG method, we assume
that the system size is even. The bond strength at the center, between
${\bf S}_{N}^{~}$ and ${\bf S}_{N+1}^{~}$,  is $J( 1 + \delta )$ when
$N$ is even and is $J( 1 - \delta )$ otherwise. The system has 2-site 
period even when $\delta = 0$, in the sense that total spin of the first $M$ site
$\sum_{i=1}^M {\bf S}_i^{~}$ alternates between integer when $M$ is 
even and half-integer when odd.

We express the ground state
wave function or its variational estimate by the notation
\begin{equation}
\Psi^{(2N)}_{~}( \sigma_1^{~} \sigma_2^{~} \ldots \sigma_N^{~} \, 
\bar \sigma^{~}_{N} \, \bar \sigma_{N-1}^{~} \ldots \bar \sigma^{~}_2 \, 
\bar \sigma^{~}_1 ) \, ,
\end{equation}
where $\sigma_i^{~} = \pm 1$ ($i \le N$) represents 
$2 S_i^{\rm Z}$ in the left half of the system, 
and where $\bar \sigma^{~}_i$ ($i \le N$) represents
$2 S_{2N+1-i}^{\rm Z}$ in the right half. We have thus divided the whole
system into the left and the right parts, according to the convention in the 
DMRG method.
Though the system described by $H^{(2N)}_{~}$ has left-right symmetry, we do not 
explicitly use it in the following formulations, in order not to loose generality.
For example, the MPS obtained by the finite system DMRG method
is not symmetric in this sense.

Let us start from the smallest case where $2N = 4$.~\cite{small} 
It is easy to numerically (or even manually) diagonalize $H^{(4)}_{~}$ to obtain the ground 
state wave function $
\Psi^{(4)}_{~}( \sigma_1^{~} \sigma_2^{~} \, \bar \sigma^{~}_2 \, \bar \sigma^{~}_1 )$.
Since we are dealing with open boundary systems, the eigenfunctions are 
always real. The density matrices for the both sides of the system
\begin{eqnarray}
\rho^{\rm L}_{~}( \sigma'_1 \sigma'_2 | \sigma_1^{~} \sigma_2^{~} ) 
\!\!\!\!\! &=& \!\!\!\!\! 
\sum_{\bar \sigma^{~}_1 \bar \sigma^{~}_2}^{~} 
\Psi^{(4)}_{~}( \sigma'_1 \sigma'_2 \, \bar \sigma^{~}_2 \, \bar \sigma^{~}_1 ) \, 
\Psi^{(4)}_{~}( \sigma_1^{~} \sigma_2^{~} \, \bar \sigma^{~}_2 \, \bar \sigma^{~}_1 ) 
\nonumber\\
\rho^{\rm R}_{~}( \bar \sigma'_1 \bar \sigma'_2 | \bar \sigma^{~}_1 \bar \sigma^{~}_2 ) 
\!\!\!\!\!  &=& \!\!\!\!\! 
\sum_{\sigma_1^{~}  \sigma_2^{~} }^{~} 
\Psi^{(4)}_{~}(  \sigma_1^{~}  \sigma_2^{~} \, \bar \sigma'_2 \, \bar \sigma'_1 ) \, 
\Psi^{(4)}_{~}( \sigma_1^{~} \sigma_2^{~} \, \bar \sigma^{~}_2 \, \bar \sigma^{~}_1 ) 
\nonumber\\
\end{eqnarray}
are therefore real symmetric. Diagonalizations of $\rho^{\rm L}_{~}$ and 
$\rho^{\rm R}_{~}$ create block spin transformations
\begin{eqnarray}
\rho^{\rm L}_{~}( \sigma'_1 \sigma'_2 | \sigma_1^{~} \sigma_2^{~} ) 
\!\!\!\!  &=&  \!\!\!\! 
\sum_{\xi_2^{~}}^{~} \lambda( \xi_2^{~} )
A_2^{~}( \sigma'_1 \sigma'_2 | \xi_2^{~} ) 
A_2^{~}( \sigma_1^{~} \sigma_2^{~} | \xi_2^{~} ) 
\nonumber\\
\rho^{\rm R}_{~}( \bar \sigma'_1 \bar \sigma'_2 | \bar \sigma^{~}_1 \bar \sigma^{~}_2 ) 
\!\!\!\!  &=& \!\!\!\! 
\sum_{\bar \xi_2^{~}}^{~} \lambda( \bar \xi_2^{~} )
B_2^{~}( \bar \sigma'_1 \bar \sigma'_2 | \bar \xi_2^{~} ) 
B_2^{~}( \bar \sigma^{~}_1 \bar \sigma^{~}_2 | \bar \xi_2^{~} ) \, ,
\nonumber\\
\end{eqnarray}
where $A_2^{~}( \sigma_1^{~} \sigma_2^{~} | \xi_2^{~} )$ and 
$B_2^{~}( \bar \sigma^{~}_1 \bar \sigma^{~}_2 | \bar \xi_2^{~} )$ 
are orthogonal matrices, respectively, which represent block spin transformations 
$\sigma_1^{~} \sigma_2^{~} \rightarrow \xi_2^{~}$ and 
$\bar \sigma^{~}_1 \bar \sigma^{~}_2 \rightarrow \bar \xi_2^{~}$. 
Thus the block spins
$\xi_2^{~}$ and $\bar \xi_2^{~}$ are 4-state variables. Applying the obtained 
(faithful) block spin transformations to $
\Psi( \sigma_1^{~} \sigma_2^{~} \, \bar \sigma^{~}_2 \, \bar \sigma^{~}_1 )$ 
we obtain the `center matrix'~\cite{McClloch}
\begin{eqnarray}
\Lambda_2^{~}( \xi_2^{~} | \bar \xi_2^{~} ) = \!\!\!\!
\sum_{ \sigma_1^{~} \sigma_2^{~} \, \bar \sigma_1^{~} \bar \sigma_2^{~} }^{~}
\!\!\!\! && \!\!\!\!\!\!\!\!\!\!\!\!\!\!
A_2^{~}( \sigma_1^{~} \sigma_2^{~} | \xi_2^{~} ) 
B_2^{~}( \bar \sigma^{~}_1 \bar \sigma^{~}_2 | \bar \xi_2^{~} ) 
\Psi( \sigma_1^{~} \sigma_2^{~} \, \bar \sigma^{~}_2 \, \bar \sigma^{~}_1 ) \, .
\nonumber\\
\end{eqnarray}
Note that the matrix $\Lambda_2^{~}( \xi_2^{~} | \bar \xi_2^{~} )$ is 
not always diagonal,
especially when we perform the diagonalizations of density matrices in Eq.~(2.4) 
independently under the condition that there is degeneracy in 
density matrix eigenvalues. It is possible to make
$\Lambda_2^{~}$ diagonal by applying singular value decomposition (SVD)
directly to $\Psi^{(4)}_{~}$, 
but we do not assume the diagonal property of the center matrices in the following.
Using the obtained matrices, we can write 
$\Psi^{(4)}_{~}$ in the form of matrix product
\begin{eqnarray}
&&\Psi^{(4)}_{~}( \sigma_1^{~} \sigma_2^{~} \, \bar \sigma^{~}_2 \, \bar \sigma^{~}_1 )
\\ 
&& = \sum_{\xi_2^{~} \bar \xi_2^{~}}^{~}
A_2^{~}( \sigma_1^{~} \sigma_2^{~} | \xi_2^{~} ) \, 
\Lambda_2^{~}( \xi_2^{~} | \bar \xi_2^{~} ) \,
B_2^{~}( \bar \sigma^{~}_1 \bar \sigma^{~}_2 | \bar \xi_2^{~} ) \, . \nonumber
\end{eqnarray}

For a while we follow the MPS construction by the infinite system DMRG method.
Then the next step is the case $2N = 6$. Applying the previously obtained 
block spin transformations $A_2^{~}$ and $B_2^{~}$ to
$H^{(6)}_{~}$ we obtain the super-block Hamiltonian ${\tilde H}^{(6)}_{~}$ that
acts to the Hilbert space spanned by $\xi_2^{~}$, $\sigma_3^{~}$, $\bar \sigma_3^{~}$, 
and $\bar \xi_2^{~}$. Diagonalizing ${\tilde H}^{(6)}_{~}$ we obtain the ground state
wave function ${\tilde \Psi}^{(6)}_{~}( \xi_2^{~} \, 
\sigma_3^{~} \, \bar \sigma_3^{~} \, \bar \xi_2^{~} )$ 
in the renormalized linear space. In the same manner as we have done 
for Eqs.~(2.3)-(2.6), we obtain the 
matrix product expression 
\begin{eqnarray}
&&{\tilde \Psi}^{(6)}_{~}( \xi_2^{~} \, \sigma_3^{~} \, \bar \sigma_3^{~} \, \bar \xi_2^{~} )
\\
&&= \sum_{\xi_3^{~} \bar \xi_3^{~}}^{~}
A_3^{~}( \xi_2^{~} \sigma_3^{~} | \xi_3^{~} ) 
\Lambda_3^{~}( \xi_3^{~} | \bar \xi_3^{~} )
B_3^{~}( \bar \xi^{~}_2 \bar \sigma^{~}_3 | \bar \xi_3^{~} ) \, , \nonumber
\end{eqnarray}
where $A_3^{~}( \xi_2^{~} \sigma_3^{~} | \xi_3^{~} )$ and
$B_3^{~}( \bar \xi^{~}_2 \bar \sigma^{~}_3 | \bar \xi_3^{~} )$ represent
block spin transformations $\xi_2^{~} \sigma_3^{~} \rightarrow \xi_3^{~}$ and
$\bar \xi^{~}_2 \bar \sigma^{~}_3 \rightarrow \bar \xi_3^{~}$. 
The dimension of the new center matrix 
$\Lambda_3^{~}( \xi_3^{~} | \bar \xi_3^{~} )$ is 8. 

\begin{figure}[h]
\begin{center}
\includegraphics[width=70mm]{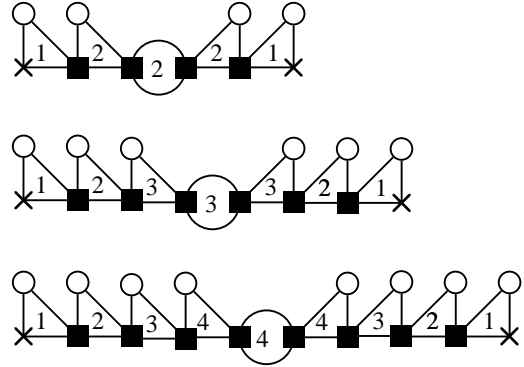}
\end{center}
\caption{\label{fig:1} Matrix product representations of ground state wave functions for 
$2N =$ 4, 6, and 8.}
\end{figure}


For convenience in the matrix product representation, let us introduce 1-state 
dummy variable $\xi_0^{~}$ and $\bar \xi_0^{~}$ at the both ends of the system. 
We put them at the both ends of the system. For example, 
$\Psi^{(4)}_{~}$ in the l.h.s. of Eq.~(2.6) can be written as 
$\Psi^{(4)}_{~}( \xi_0^{~} \sigma_1^{~} 
\sigma_2^{~} \, \bar \sigma_2^{~} \, \bar \sigma_1^{~} \, \bar \xi_0^{~} )$; 
if we neglect the dummy variable the original form of $\Psi^{(4)}_{~}$ in Eq.~(2.6) is
recovered. 
Then how does the r.h.s of Eq.~(2.6) look like? In order to answer this question
we also introduce two state block spin variables $\xi_1^{~}$ and $\bar \xi_1^{~}$,
respectively, which is always the same as $\sigma_1^{~}$ and $\bar \sigma_1^{~}$.
Using these variables we define the boundary orthogonal matrices
\begin{eqnarray}
A_1^{~}( \xi_0^{~} \sigma_1^{~} | \xi_1^{~} ) \!\!\!\! &=& \!\!\!\!
\delta( \sigma_1^{~} | \xi_1^{~} ) \nonumber\\
B_1^{~}( \bar \xi_0^{~} \bar \sigma_1^{~} | \bar \xi_1^{~} ) \!\!\!\! &=& \!\!\!\!
\delta( \bar \sigma_1^{~} | \bar \xi_1^{~} ) \, ,
\end{eqnarray}
where $\delta( a | b )$ represents Kronecker's delta $\delta_{ab}^{~}$.
With the help of these boundary orthogonal matrices, we can express $\Psi^{(4)}_{~}$ and 
$\Psi^{(6)}_{~}$ in the matrix product form
\begin{eqnarray}
\Psi^{(4)}_{~} \!\!\!\! &=& \!\!\!\! A_1^{~} A_2^{~} \Lambda_2^{~} B_2^{\dagger} B_1^{\dagger} 
\nonumber\\
\Psi^{(6)}_{~} \!\!\!\! &=& \!\!\!\! A_1^{~} A_2^{~} A_3^{~} \Lambda_3^{~} 
B_3^{\dagger} B_2^{\dagger} B_1^{\dagger} \, ,
\end{eqnarray}
where we have changed the variables of $A_2^{~}$ and $B_2^{~}$ as
$A_2^{~}( \xi_1^{~} \sigma_2^{~} | \xi_2^{~} )$ and 
$B_2^{~}( \bar \xi_1^{~} \bar \sigma_2^{~} | \bar \xi_2^{~} )$, respectively.
In equation (2.9) we regard block spin variables $\xi_i^{~}$ and $\bar \xi_i^{~}$ 
as the matrix index and
take their configuration sum, leaving the raw spin variables $\sigma_i^{~}$ and
$\bar \sigma_i^{~}$. It might be
better to regard $A_i^{~}$ and $B_i^{~}$ as 3-leg tensors, and r.h.s. of
the above equation as tensor products.~\cite{int} 

It is straight forward to extend the matrix product expression of the ground 
state wave function to arbitrary system size
\begin{eqnarray}
\Psi^{(2N)}_{~} \!\!\!\! &=& \!\!\!\! A_1^{~} A_2^{~} \ldots A_N^{~} 
\Lambda_N^{~} 
B_N^{\dagger} \ldots B_2^{\dagger} B_1^{\dagger} \\
\!\!\!\! &=& \!\!\!\!
A_1^{~} A_2^{~} \ldots A_{N-1}^{~} 
{\tilde \Psi}^{(2N)}_{~}
B_{N-1}^{\dagger} \ldots B_2^{\dagger} B_1^{\dagger}
\, , \nonumber
\end{eqnarray}
where configuration sum is taken for all the block spin variables,
and where ${\tilde \Psi}^{(2N)}_{~} = A_N^{~}  \Lambda_N^{~} B_N^{\dagger}$.
Figure 1 shows the graphical representation of $\Psi^{(2N)}_{~}$ for $2N = 4, 6,$ and $8$, 
where cross marks represent the dummy variables $\xi_0^{~}$ and $\bar \xi_0^{~}$ at 
the both ends, black squares the block spin variables, and circles the raw spin variables.
From the computational view point, it is impossible to keep all the degrees of freedom
in block spin transformation for arbitrary large system size, therefore the number of
state of the block spin variables $\xi_i^{~}$ and $\bar \xi_i^{~}$ are restricted at 
most $m$ states. When there is a cut off in this sense, the r.h.s. of Eq.~(2.10) is
a variational approximation for the l.h.s. For example, 
 $A_N^{~}  \Lambda_N^{~} B_N^{\dagger}$ is an approximation of
 ${\tilde \Psi}^{(2N)}_{~}$ when the matrix dimension of $\Lambda_N^{~}$ is restricted.

Though we have created the matrix product
wave function in Eq.~(2.10) by way of the infinite system DMRG method, we do not
restrict ourselves about the way of creation of MPS in the following formulation. For 
example, we also deal MPS obtained by the finite system DMRG method, where 
the sweeping is stopped at the center of the system. Strictly speaking, 
the matrices $A_i^{~}$ and $B_i^{~}$ determined by the finite system DMRG method
is dependent to the system size $2N$, there fore we have to put the system size
to the matrix labels as $A_i^{(2N)}$ and $B_i^{(2N)}$ for distinction. But the notation
is rather complicated, and therefore we drop the label $(2N)$ in the following equations.

Let us observe the renormalized wave function ${\tilde \Psi}^{(2N)}_{~}$, which
corresponds to the lowest energy state of the super-block Hamiltonian ${\tilde H}^{(2N)}_{~}$. 
It is possible to obtain ${\tilde \Psi}^{(2N)}_{~}$ applying block spin
transformations $A_1^{~} \ldots A_{N-1}^{~}$ and 
$B_1^{~} \ldots B_{N-1}^{~}$ successively to $\Psi^{(2N)}_{~}$ as
\begin{equation}
{\tilde \Psi}^{(2N)}_{~} = \!\!\!\! \sum_{\sigma_1^{~} \ldots \sigma_{N-1}^{~} \,
\bar \sigma_{N-1}^{~}  \ldots \bar \sigma_1^{~} }^{~} \!\!\!\! 
A_{N-1}^{\dagger} \ldots A_{1}^{\dagger} \Psi^{(2N)}_{~} B_{1}^{~} \ldots B_{N-1}^{~} \, ,
\end{equation}
where we have identified the wave function $\Psi^{(2N)}_{~}$ as a 3-leg tensor,
which has (dummy) matrix indices $\xi_0^{~}$ and $\bar \xi_0^{~}$ in addition to the 
row spin variables $\{ \sigma \} = \sigma_1^{~} \ldots \sigma_{N}^{~} \, 
\bar \sigma_{N}^{~}  \ldots \bar \sigma_1^{~}$. 

\section{Wave Function Renormalization}

Suppose we have matrix product expressions for 
$\Psi^{(4)}_{~}$ and $\Psi^{(6)}_{~}$ in Eq.~(2.9), and need to obtain that of
$\Psi^{(8)}_{~}$. This need is fulfilled if we diagonalize the 
Hamiltonian $H^{(8)}_{~}$ via eigen solver such as the Lanczos 
method. Under the situation it is important to prepare
a good trial (or initial) wave function for the numerical 
diagonalization process. An answer to this problem can be obtained 
from observation on the bare Hamiltonians $H^{(4)}_{~}$ and $H^{(8)}_{~}$.
Since these two Hamiltonians has the same bond strength at the center
of the system, $\Psi^{(4)}_{~}$ can be used as a trial (or variational) wave function for
$H^{(8)}_{~}$ if we put two additional spins to the both ends. 
This construction is represented as
\begin{equation}
\Psi_{\rm trial}^{(8)}( \sigma_1^{~}  \sigma_2^{~} \sigma_3^{~} \sigma_4^{~} \, 
\bar \sigma_4^{~} \, \bar \sigma_3^{~} \, \bar \sigma_2^{~} \,  \bar \sigma_1^{~} ) = 
\Psi^{(4)}_{~}( \sigma_3^{~} \sigma_4^{~} \, \bar \sigma_4^{~} \, \bar \sigma_3^{~} ) \, 
\end{equation}
apart from the normalization factor,
where the trial wave function $\Psi_{\rm trial}^{(8)}$ is not dependent to
$\sigma_1^{~}$, $\sigma_2^{~}$, $\bar \sigma_2^{~}$, and $\bar \sigma_1^{~}$.
Such a construction of trial wave function can be generalized to arbitrary system size $2N$,
where $\Psi^{(2N+2)}_{\rm trial}$ is obtained from $\Psi^{(2N-2)}_{~}$. Since this is
a rough estimation, one has to improve the trial wave function afterward.

\begin{figure}[h]
\begin{center}
\includegraphics[width=70mm]{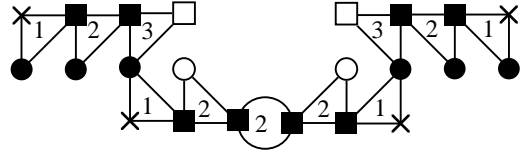}
\end{center}
\caption{\label{fig:2} Graphical representation of the trial wave function estimation for $2N = 8$.}
\end{figure}


Consider the way of expressing the wave function estimation in Eq.~(3.1) in 
the renormalized subspace. 
Applying block spin transformations, which are already obtained up to $2N = 6$, to
the estimated wave function $\Psi_{\rm trial}^{(8)}( \sigma_1^{~}  \sigma_2^{~} 
\sigma_3^{~} \sigma_4^{~} \, \bar \sigma_4^{~} \, \bar \sigma_3^{~} \, \bar 
\sigma_2^{~} \,  \bar \sigma_1^{~} )$, we obtain
the  renormalized form of the trial wave function 
\begin{equation}
{\tilde \Psi}_{\rm trial}^{(8)}( \xi_3^{~} \sigma_4^{~} \, \bar \sigma_4^{~} \, 
\bar \xi_3^{~} ) = \!\!\!\!
\sum_{\sigma_1^{~} \sigma_2^{~} \sigma_3^{~} \bar \sigma_3^{~} \bar \sigma_2^{~}
\bar \sigma_1^{~}}^{~} \!\!\!\!
A_3^{\dagger} A_2^{\dagger} A_1^{\dagger} \, \Psi_{\rm trial}^{(8)} \, 
B_1^{~} B_2^{~} B_3^{~} 
\end{equation}
as we have done in Eq.~(2.11).
Figure 2 shows the graphical representation of the above wave function 
renormalization process applied to $\Psi_{\rm trial}^{(8)}$, where we
draw $\Psi^{(4)}_{~}$ by its matrix product expression. 
In order to write Eq.~(3.2) more transparently, we introduce dummy 
matrices
\begin{eqnarray}
A_{0}^{~}( \xi_{-1}^{~} \sigma_2^{~} | \xi_0^{~} ) ~~ &=& 1 / \sqrt{2} \nonumber\\
A_{-1}^{~}( \xi_{-2}^{~} \sigma_1^{~} | \xi_{-1}^{~} ) &=& 1 / \sqrt{2} \nonumber\\
B_{0}^{~}( \bar \xi_{-1}^{~} \bar \sigma_2^{~} | \bar \xi_0^{~} ) 
~~ &=& 1 / \sqrt{2} \nonumber\\
B_{-1}^{~}( \bar \xi_{-2}^{~} \bar \sigma_1^{~} | \bar \xi_{-1}^{~} ) &=& 1 / \sqrt{2} \, ,
\end{eqnarray}
where $\xi_{-2}^{~}$, $\xi_{-1}^{~}$, $\xi_{0}^{~}$, 
$\bar \xi_{-2}^{~}$, $\bar \xi_{-1}^{~}$, and $\bar \xi_{0}^{~}$ are 1-state
dummy variables. Then the $\Psi_{\rm trial}^{(8)}$ in Eq.~(3.1) can be written as
\begin{eqnarray}
\Psi_{\rm trial}^{(8)} 
&=& 
A_{-1}^{~} A_0^{~} A_1^{~} A_2^{~} \Lambda_2^{~}
B_2^{\dagger} B_1^{\dagger} B_0^{\dagger} B_{-1}^{\dagger} \nonumber\\
&=&
A_{-1}^{~} A_0^{~} A_1^{~} {\tilde \Psi}^{(4)}_{~}
B_1^{\dagger} B_0^{\dagger} B_{-1}^{\dagger} \, ,
\end{eqnarray}
where ${\tilde \Psi}^{(4)}_{~}( \xi_1^{~} \sigma_4^{~} \, \bar \sigma_4^{~} \, \bar \xi_1^{~} )$ 
is nothing but ${\Psi}^{(4)}_{~}( \sigma_1^{~} \sigma_4^{~} \, 
\bar \sigma_4^{~} \, \bar \sigma_1^{~} )$ since $\xi_1^{~} = \sigma_1^{~}$ and 
$\bar \xi_1^{~} = \bar \sigma_1^{~}$ by the definition of $A_1^{~}$ and $B_1^{~}$
in Eq.~(2.8).
It should be noted that in Eqs.~(3.3) and (3.4) the matrix labels are shifted by 2,
in the sense that $A_1^{~}$, $A_2^{~}$, $B_1^{~}$, and $B_2^{~}$, respectively,
contains $\sigma_3^{~}$, $\sigma_4^{~}$, $\bar \sigma_3^{~}$, and
$\bar \sigma_4^{~}$.
Substituting Eq.~(3.4) into Eq.~(3.2) we obtain
\begin{eqnarray}
{\tilde \Psi}_{\rm trial}^{(8)} \!\!\!\! &=& \!\!\!\!\!\!\!\!\!\! 
\sum_{\sigma_1^{~} \sigma_2^{~} \sigma_3^{~} \, 
\bar \sigma_3^{~} \bar \sigma_2^{~} \bar \sigma_1^{~}}^{~} \!\!\!\!\!\!
A_3^{\dagger} A_2^{\dagger} A_1^{\dagger} 
A_{-1}^{~} A_0^{~} A_1^{~} {\tilde \Psi}^{(4)}_{~} 
B_1^{\dagger} B_0^{\dagger} B_{-1}^{\dagger}
B_1^{~} B_2^{~} B_3^{~} \nonumber\\
&=& \sum_{\xi^{~}_1 \bar \xi^{~}_1}^{~}
L_3^{~}( \xi_3^{~} | \xi^{~}_1 ) \,
{\Psi}^{(4)}_{~}( \xi^{~}_1 \sigma_4^{~} \, \bar \sigma_4^{~} \, \bar \xi^{~}_1 ) \,
R_3^{~}( \bar \xi_3^{~} | \bar \xi^{~}_1 ) \nonumber\\
&=& 
L_3^{~} \,  {\Psi}^{(4)}_{~} R^{\dagger}_{3} \, ,
\end{eqnarray}
where the matrix $L_3^{~}( \xi_3^{~} | \xi_1 )$ is defined as
\begin{eqnarray}
L_3^{~}( \xi_3^{~} | \xi_1^{~} ) 
\!\!\!\! &=& \!\!\!\!
\sum_{\sigma_1^{~} \sigma_2^{~} \sigma_3^{~}}^{~}
A_3^{\dagger} A_2^{\dagger} A_1^{\dagger} 
A_{-1}^{~} A_0^{~} A_1^{~} \\
&=& \!\!\!\!\!\!
\sum_{\sigma_1^{~} \sigma_2^{~} \sigma_3^{~} \xi_2^{~}}^{~}
A_2^{~}( \sigma_1^{~} \sigma_2^{~} | \xi_2^{~} ) \, 
A_3^{~}( \xi_2^{~} \sigma_3^{~} | \xi_3^{~} )
A_1^{~}( \xi_0^{~} \sigma_3^{~} | \xi_1^{~} )
\nonumber
\end{eqnarray}
and $R_3^{~}( \bar \xi_3^{~} | \bar \xi^{~}_1 )$ is in the same manner
\begin{eqnarray}
R_3^{~}( \bar \xi_3^{~} | \bar \xi_1^{~} ) 
\!\!\!\! &=& \!\!\!\!
\sum_{\bar \sigma_1^{~} \bar \sigma_2^{~} \bar \sigma_3^{~} }^{~}
B_3^{\dagger} B_2^{\dagger} B_1^{\dagger} 
B_{-1}^{~} B_0^{~} B_1^{~} \\
&=& \!\!\!\!
\sum_{\bar \sigma_1^{~} \bar \sigma_2^{~} \bar \sigma_3^{~}  \xi_2^{~}}^{~}
B_2^{~}( \bar \sigma_1^{~} \bar \sigma_2^{~} | \bar \xi_2^{~} ) \, 
B_3^{~}( \bar \xi_2^{~} \bar \sigma_3^{~} | \bar \xi_3^{~} ) 
B_1^{~}( \bar \xi_1^{~} \bar \sigma_1^{~} | \bar \xi_1^{~} )
\nonumber
\end{eqnarray}
Figure 3 shows the graphical representation of $L$ and $R$, the
matrices which have a function of `adjusting' the dimension of block spin variables.

\begin{figure}[h]
\begin{center}
\includegraphics[width=75mm]{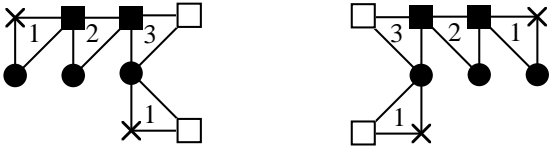}
\end{center}
\caption{\label{fig:3} Graphical representations of $L$ (left) and $R$ (right) in Eq.~(3.6).}
\end{figure}


As $\Psi_{\rm trial}^{(8)}$ can be used as a trial wave function for $H^{(8)}_{~}$, 
${\tilde \Psi}_{\rm trial}^{(8)}$ thus obtained 
would be a trial wave function for the super-block Hamiltonian ${\tilde H}^{(8)}_{~}$. 
Now we can perform Lanczos diagonalization of ${\tilde H}^{(8)}_{~}$ rapidly
starting from ${\tilde \Psi}_{\rm trial}^{(8)}$. Let us assume that we obtain the 
improved ${\tilde \Psi}^{(8)}_{~}( \xi_3^{~} \sigma_4^{~} \, \bar \sigma_4^{~} \,
\bar \xi_3^{~} )$ in this way. From the calculated ${\tilde \Psi}^{(8)}_{~}$ we obtain
$A_4^{~}( \xi_3^{~} \sigma_4^{~} | \xi_4^{~} )$,
$B_4^{~}( \bar \xi_3^{~} \bar \sigma_4^{~} | \bar \xi_4^{~} )$, and
$\Lambda_4^{~}( \xi_4^{~} | \bar \xi_4^{~} )$ as we have done in previous steps. 

We can extend the way of initial wave function estimation to the case
$2N = 10$, where 
${\tilde \Psi}_{\rm trial}^{(10)}( \xi_4^{~} \sigma_5^{~} \, \bar \sigma_5^{~} 
\bar \xi_4^{~} )$ is required. This prediction is performed as 
\begin{eqnarray}
{\tilde \Psi}_{\rm trial}^{(10)} \!\!\!\! &=& \!\!\!\!
L_4^{~} {\tilde \Psi}^{(6)}_{~} R_4^{\dagger} \\
&=& \!\!\!\!
\sum_{\xi_2^{~} \bar \xi_2^{~}}^{~}
L_4^{~}( \xi_4^{~} | \xi_2^{~} ) {\tilde \Psi}^{(6)}_{~}( \xi_2^{~} \sigma_5^{~} \, \bar \sigma_5^{~} \, 
\bar \xi_2^{~} ) R_4^{~}( \bar \xi_4^{~} | \bar \xi_2^{~} ) \, , \nonumber
\end{eqnarray}
where $L_4^{~}$ and $R_4^{~}$ are defined as follows
\begin{eqnarray}
L_4^{~}( \xi_4^{~} | \xi_2^{~} ) \!\!\!\! &=& \!\!\!\!
\sum_{\sigma_4^{~}}^{~} A_4^{\dagger} L_3^{~} A_2^{~} \nonumber\\
&=& \!\!\!\!
\sum_{ \xi_3^{~} \xi_1^{~} \sigma_4^{~}}^{~} 
A_4^{~}( \xi_3^{~} \sigma_4^{~} | \xi_4^{~} )
L_3^{~}( \xi_3^{~} | \xi_1^{~} )
A_2^{~}( \xi_1^{~} \sigma_4^{~} | \xi_2^{~} )
\nonumber\\
R_4^{~}( \bar \xi_4^{~} | \bar \xi_2^{~} ) \!\!\!\! &=& \!\!\!\!
\sum_{\bar \sigma_4^{~}}^{~} B_4^{\dagger} R_3^{~} B_2^{~} \nonumber\\
&=& \!\!\!\!
\sum_{ \bar \xi_3^{~} \bar \xi_1^{~} \bar \sigma_4^{~}}^{~} 
B_4^{~}( \bar \xi_3^{~} \bar \sigma_4^{~} | \bar \xi_4^{~} )
R_3^{~}( \bar \xi_3^{~} | \bar \xi_1^{~} )
B_2^{~}( \bar \xi_1^{~} \bar \sigma_4^{~} | \bar \xi_2^{~} ) \, .
\nonumber\\
\end{eqnarray}
These recursive relations in $L_N^{~}$ and $R_N^{~}$ in the above equation
was first obtained empirically and has been used for the numerical study by use 
of the PWFRG method when it is applied to $S = 1/2$ quantum spin 
chains.~\cite{Hieida,Hagiwara,Okunishi,Narumi,Yoshikawa}
Figure 4 show the graphical representation of the relation between
$L_3^{~}$ and $L_4^{~}$, and also between $R_3^{~}$ and $R_4^{~}$.
The process of wave function estimation is drawn in Fig.~5.

\begin{figure}[h]
\begin{center}
\includegraphics[width=35mm]{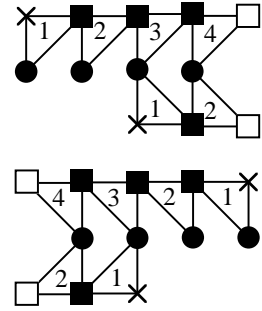}
\end{center}
\caption{\label{fig:4} Construction of $L_4^{~}$ (upper) and $R_4^{~}$ (lower) in Eq.~(3.9).}
\end{figure}


\begin{figure}[h]
\begin{center}
\includegraphics[width=75mm]{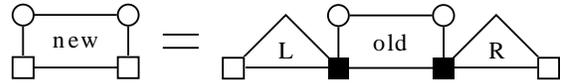}
\end{center}
\caption{\label{fig:5} Graphical representation of the wave function estimation.}
\end{figure}


It is straight forward to extend the relation in Eqs.~(3.8) and (3.9) 
for arbitrary system size. This is the way of wave function estimation,
which we call as the 2-site shift PWFRG method. We can obtain 
${\tilde \Psi}^{(2N + 2)}_{\rm trial}$ if we have matrix product expressions
of both $\Psi^{(2N)}_{~}$ and $\Psi^{(2N-2)}_{~}$. 
The wave function estimation in Eq.~(3.8) is performed using $\Psi^{(2N-2)}_{~}$
directly, instead of its matrix product decomposition 
$A_{N-1}^{~} \Lambda_{N-1}^{~} B_{N-1}^{\dagger}$ where basis 
state restriction is imposed on $\Lambda_{N-1}^{~}$. Thus the
estimation becomes exact in the thermodynamic limit, where the matrix product 
wave function is position independent. In this sense the way of estimation
explained here is better than the estimation in the previous formulation
of the PWFRG method,~\cite{PWFRG,PWFRG2} which uses truncated 
$A_{N-1}^{~} \Lambda_{N-1}^{~} B_{N-1}^{\dagger}$, when the
infinite system DMRG method is nearly converged.~\cite{commun} 

It should be
noted that there is no need that $\Psi^{(2N)}_{~}$ and $\Psi^{(2N-2)}_{~}$
have the same matrices in common. For example, the estimation 
for ${\tilde \Psi}^{(2N + 2)}_{\rm trial}$
can be performed, if we have optimized ground states for $(2N-2)$- and $2N$-site
systems independently by use of the finite system DMRG method. In such a case 
the matrices $A_i^{~}$ and $B_i^{~}$ becomes system size dependent,
as we have seen at the end of the last section. The definition of $L_i^{~}$ and $R_i^{~}$ 
should be modified according to the dependence, where the extension is 
straight forward.

\section{Convergence to the Thermodynamic Limit}

The estimated wave function 
\begin{eqnarray}
&&\Psi^{(2N+2)}_{\rm trial}( \sigma_1^{~} \ldots \sigma_{N+1}^{~} \, 
\bar \sigma_{N+1}^{~} \ldots \bar \sigma_1^{~} ) \nonumber\\
&&= 
\Psi^{(2N-2)}_{~}( \sigma_3^{~} \ldots \sigma_{N-1}^{~} \, 
\bar \sigma_{N-1}^{~} \ldots \sigma_1^{~} ) \, 
\end{eqnarray}
is normally not accurate enough, 
since the estimated wave function is independent of 2 spins at each end 
of the extended system. Therefore the estimated renormalized wave function
\begin{equation}
{\tilde \Psi}_{\rm trial}^{(2N+2)} = L_{N}^{~} {\tilde \Psi}^{(2N-2)}_{~}
R_{N}^{\dagger}
\end{equation}
might not be a good starting point for the Lanczos diagonalization of 
${\tilde H}^{(2N+2)}_{~}$. Let us check the efficiency in the estimation
quantitatively by use of the fidelity error~\cite{McClloch}
\begin{eqnarray}
1 - 
\sum_{ \xi_{N}^{~} \sigma_{N+1}^{~} \, \bar \sigma_{N+1}^{~} \, \bar \xi_{N}^{~} }^{~}
\!\!\!\!\!\!\!\! && 
\!\!\!\!\!\!\!\! {\tilde \Psi}_{\rm trial}^{(2N+2)}( \xi_{N}^{~} \sigma_{N+1}^{~} \, 
\bar \sigma_{N+1}^{~} \, \bar \xi_{N}^{~} ) \nonumber\\
&& \!\!\!\!\!\!\!\! {\tilde \Psi}_{~}^{(2N+2)}( \xi_{N}^{~} \sigma_{N+1}^{~} \, 
\bar \sigma_{N+1}^{~} \, \bar \xi_{N}^{~} ) \, ,
\end{eqnarray}
between normalized ${\tilde \Psi}_{\rm trial}^{(2N+2)}$ and ${\tilde \Psi}_{~}^{(2N+2)}$.
We observe the error when the wave function estimation is implemented in the
infinite system DMRG method. The computational algorithm we have used
for this check is as follows.
\vskip 2mm
\begin{itemize}
\item[(a)] Diagonalize $H^{(4)}_{~}$ and obtain $\Psi^{(4)}_{~} = {\tilde \Psi}^{(4)}_{~}$,
$A_2^{~}$, and $B_2^{~}$.
\item[(b)] Create ${\tilde H}^{(6)}_{~}$ by applying $A_2^{~}$ and $B_2^{~}$ to
$H^{(6)}_{~}$. Diagonalize ${\tilde H}^{(6)}_{~}$ and obtain ${\tilde \Psi}^{(6)}_{~}$,
$A_3^{~}$, and $B_3^{~}$.
\item[(c)] Contracting $A_3^{~}$ and $B_3^{~}$ as Eqs.~(3.6) and (3.7), respectively,
to obtain $L_3^{~}$ and $R_3^{~}$. Set $N = 3$.
\item[(d)] Obtain ${\tilde \Psi}^{(2N+2)}_{\rm trial}$ by applying $L_N^{~}$
and  $R_N^{\dagger}$ to ${\tilde \Psi}^{(2N-2)}_{~}$
 as shown in Eqs.~(3.5) and (3.8).
\item[(e)] Create the super-block Hamiltonian ${\tilde H}^{(2N+2)}_{~}$.
\item[(f)] Obtain minimum eigenvalue of ${\tilde H}^{(2N+2)}_{~}$ and corresponding
wave function, starting from ${\tilde \Psi}^{(2N+2)}_{\rm trial}$. 
\item[(g)] Obtain $A_{N+1}^{~}$ and $B_{N+1}^{~}$. By use of these 
transformations, create $L_{N+1}^{~} = 
\sum_{\sigma_{N+1}^{~}}^{~} A_{N+1}^{\dagger} L_N^{~} A_{N-1}^{~}$ and
$R_{N+1}^{~} = 
\sum_{\bar \sigma_{N+1}^{~}}^{~} B_{N+1}^{\dagger} R_N^{~} B_{N-1}^{~}$ 
as Eq.~(3.9)
\item[(h)] Set $N + 1 \rightarrow N$ and go to the step (d).
\end{itemize}
\vskip 2mm

\begin{figure}[h]
\begin{center}
\includegraphics[width=75mm]{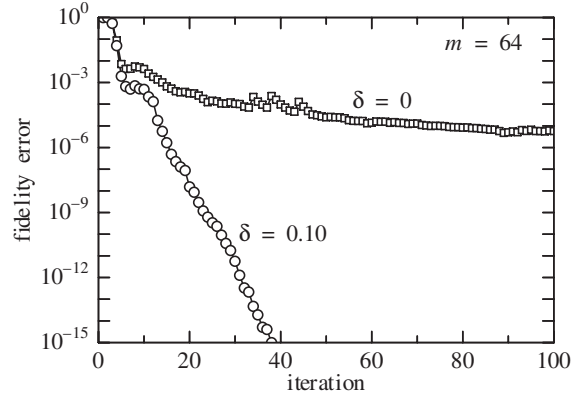}
\end{center}
\caption{\label{fig:6} The fidelity error in Eq.~(4.3) calculated for the 
uniform $S = 1/2$ Heisenberg spin chain $\delta = 0$ and the 
dimerized one $\delta = 0.1$.}
\end{figure}


Figure 6 shows the fidelity error of the $S = 1/2$ Heisenberg spin
chain with respect to the system size $2N$ 
when $\delta = 0$ and $\delta = 0.1$ under the condition $m = 64$. 
In both cases the error decreases with the system size, and the decay
is more rapid when $\delta = 0.1$ than $\delta = 0$. The
behavior can be explained by the fact that the effect of system boundary becomes
weak in large size systems, and the center of the system is `effectively 
decoupled' from the system boundary if the system size exceeds 
several time larger than the correlation length.
As we have stated in the last section, the estimation becomes exact 
${\tilde \Psi}_{\rm trial}^{(2N+2)} = {\tilde \Psi}_{~}^{(2N+2)}$
in the thermodynamic limit $N \rightarrow \infty$, where the 
fidelity error becomes zero.

It has been known that the DMRG method applied to gapless systems
introduces an artificial correlation
length, as an effect of basis state restriction to $m$. Therefore the
convergence of the fidelity error with respect to the system size $2N$ is 
slow exponential when the system size $2N$ is sufficiently large. 
In such a case, it is better to increase the system size as fast as possible.
The recursion relation
\begin{eqnarray}
L_{N+1}^{~} &=& \sum_{\sigma_{N+1}^{~}}^{~} 
A_{N+1}^{\dagger} L_N^{~} A_{N-1}^{~} \nonumber\\
R_{N+1}^{~} &=& \sum_{\sigma_{N+1}^{~}}^{~} 
B_{N+1}^{\dagger} R_N^{~} B_{N-1}^{~} 
\end{eqnarray}
can be regarded as linear transformations to $L_{N}^{~}$ and $R_{N}^{~}$, which
have their fixed points in the limit $N \rightarrow \infty$. If the number of 
block spin states does not change during this extension process, 
one can estimate these fixed point easily. But the number of states of the block spin
variables are not always the same. A way of overcoming this difficulty
is to modify the diagonalization step (f) as follows.~\cite{Hieida,PWFRG2}
~\\
\begin{itemize}
\item[(f')] Improve the estimated wave function ${\tilde \Psi}_{\rm trial}^{(2N+2)}$
by applying the Lanczos step only once.  Use the `improved' wave function 
${\tilde \Psi}_{\rm improved}^{(2N+2)}$ for the succeeding processes.
\end{itemize}
~\\

When the system is gapless, the efficiency of the PWFRG method decreases.
Quite recently McClloch reported a new estimation scheme, which works even for
gapless systems. Let us observe his method from the view angle of 
the wave function renormalization.
The starting point is to interpret the wave function as a matrix
\begin{equation}
\Psi^{(2N)}_{~}( \sigma_1^{~} \ldots \sigma_N^{~} | 
\, \bar \sigma^{~}_N \ldots \bar \sigma_1^{~}  ) \, .
\end{equation}
This `wave function matrix' satisfies the identity relation
\begin{equation}
\Psi^{(2N)}_{~} \left( \Psi^{(2N)}_{~} \right)^{-1}_{~} \Psi^{(2N)}_{~} = \Psi^{(2N)}_{~} \, ,
\end{equation}
where $\left( \Psi^{(2N)}_{~} \right)^{-1}_{~}$ is matrix inverse of $\Psi^{(2N)}_{~}$.
McClloch's way of wave function estimation is obtained by 
decreasing the system size of this inverse matrix by 2
\begin{equation}
\Phi_{\rm L}^{(2N)} \left( \Psi^{(2N-2)}_{~} \right)^{-1}_{~} \Phi_{\rm R}^{(2N)} 
= \Psi^{(2N+2)}_{\rm trial} \, ,
\end{equation}
where we $\Phi_{\rm L}^{(2N)}$ and $\Phi_{\rm R}^{(2N)}$ are rectangular 
matrices
\begin{eqnarray}
&&\Phi_{\rm L}^{(2N)} = 
\Psi_{~}^{(2N)}( \sigma_1^{~} \ldots \sigma_N^{~} \, \bar \sigma^{~}_N | \,
\bar \sigma^{~}_{N-1} \ldots \bar \sigma_1^{~}  ) \nonumber\\
&&\Phi_{\rm R}^{(2N)}  = 
\Psi_{~}^{(2N)}( \sigma_1^{~} \ldots \sigma^{~}_{N-1} | \sigma_N^{~} \, 
\bar \sigma^{~}_N   \ldots \bar \sigma_1^{~}  ) 
\end{eqnarray}
obtained by shifting the left-right division of the system by 1 site.
This construction is similar to the extension of corner transfer matrix,
which has been applied to two-dimensional classical lattice models.~\cite{Baxter3}

It is easy to see that McClloch's  way of wave function estimation can be
performed by use of $\Psi^{(2N-2)}_{~}$ and $\Psi^{(2N)}_{~}$ that are created
independently by the finite system DMRG method. Let us express
$\Psi^{(2N-2)}_{~}$ and $\Psi^{(2N)}_{~}$ as
\begin{eqnarray}
\Psi^{(2N-2)}_{~} \!\!\!\!  &=& \!\!\!\! 
A_1^{~} \ldots A_{N-1}^{~} \Lambda_{N-1}^{~}
B_{N-1}^{\dagger} \ldots B_1^{\dagger}
\nonumber\\
\Psi^{(2N)}_{~} \!\!\!\!  &=& \!\!\!\! 
{A'}_1 \ldots {A'}_{N} {\Lambda'}_{N}^{~}
{B'}_{N}^{\dagger} \ldots {B'}_1^{\dagger} \, ,
\end{eqnarray}
where $A_i^{~} = A_i^{(2N-2)}$ and $B_i^{~} = B_i^{(2N-2)}$ are not always the same as
${A'}_i^{~} = {A}_i^{(2N)}$ and ${B'}_i^{~} = {B}_i^{(2N)}$, respectively. 
Substituting these matrix product wave functions to Eq.~(4.7) we obtain
\begin{eqnarray}
\Psi_{\rm trial}^{(2N+2)} = \!\!\!\!  && \!\!\!\! 
{A'}_1^{~} \ldots {A'}_{N}^{~} {\Lambda'}_{N}^{~}  {B'}_{N}^{\dagger} R^{\dagger}_{~}
\left( \Lambda_{N-1}^{~} \right)^{-1}_{~} L \nonumber\\
&&{A'}_{N} {\Lambda'}_{N}^{~} {B'}_{N}^{\dagger} \ldots {B'}_1^{\dagger} \, ,
\end{eqnarray}
where the matrices $L$ and $R$ are defined as follows
\begin{eqnarray}
L \!\!\!\! &=& \!\!\!\! \sum_{\sigma_1^{~} \ldots \sigma_{N-1}^{~}}^{~}
{A}_{N-1}^{\dagger} \ldots {A}_1^{\dagger} {A'}_1^{~} \ldots {A'}_{N-1}^{~}
\nonumber\\
R \!\!\!\! &=& \!\!\!\! \sum_{\bar \sigma_1^{~} \ldots \, \bar \sigma_{N-1}^{~}}^{~}
{B}_{N-1}^{\dagger} \ldots {B}_1^{\dagger} {B'}_1^{~} \ldots {B'}_{N-1}^{~} \, .
\end{eqnarray}
%
Note that the matrices $L$ and $R$ becomes identity ones 
when both $\Psi^{(2N-2)}_{~}$ and $\Psi^{(2N)}_{~}$ are created succeedingly  
by the infinite system DMRG algorithm.

\section{Conclusions}

We have formulated a way of applying the PWFRG method for quantum spin systems
which have 2-site modulation. In order to estimate the initial wave function, we shift the 
application of renormalization group transformation to the wave function by 2 lattice cites. 
As a result, we obtain a recursive relation among renormalized wave functions.
Numerical efficiency of the wave function estimation is confirmed when the method is 
applied to the dimerized $S = 1/2$ Heisenberg spin chain. We give an interpretation
to McClloch's way of wave function estimation, from the view point of wave function 
renormalization.

We thank to I. McClloch for valuable comments and discussions. 
H.~U thank to Dr. Okunishi for helpful comments on DMRG and 
continuous encouragement. T.~N is partially supported by a Grant-in-Aid 
for Scientific Research from the Ministry 
of Education, Science, Sports and Culture.


\begin{thebibliography}{99}
\bibitem{KW} H.A.~Kramers and G.H.~Wannier: Phys. Rev. {\bf 60} (1941) 263.
\bibitem{Kikuchi} R.~Kikuchi: Phys. Rev. {\bf 81} (1951) 988.
\bibitem{Baxter1} R.J.~Baxter: J. Math. Phys {\bf 9} (1968) 650.
\bibitem{Baxter2} R.J.~Baxter: J. Stat. Phys. {\bf 19} (1978) 461.
\bibitem{Baxter3} R.J.~Baxter: {\it Exactly Solved Models in Statistical
Mechanics}, Academic Press, London (1982).
\bibitem{NB} N.P.~Nightingale and H.W.~Bl\"ote: Phys. Rev. B {\bf 33} (1986) 659.
\bibitem{AKLT} I.~Affleck, T.~Kennedy, E.H.~Lieb, and H.~Tasaki: Phys.
Rev. Lett. {\bf 59} (1987) 799.
\bibitem{Fannes1} M.~Fannes, B.~Nachtergale  and R.~F.~Werner: Europhys. Lett. 
{\bf 10} (1989) 633.
\bibitem{Fannes2} M.~Fannes, B.~Nachtergale  and R.~F.~Werner: Commun. Math. 
Phys. {\bf 144} (1992)  443.
\bibitem{Fannes3} M.~Fannes, B.~Nachtergale  and R.~F.~Werner: Commun. Math. Phys. {\bf 174}  (1995) 477.
\bibitem{Klumper1} A.~Kl\"umper, A.~Schadschneider,  and J.~Zittartz: 
Z. Phys. B {\bf 87} (1992) 281.
\bibitem{Klumper2} H.~Niggemann, A.~Kl\"umper, and J.~Zittartz: 
Z. Phys. B {\bf 104} (1997) 103.
\bibitem{DMRG} S.~R.~White: Phys. Rev. Lett. {\bf 69} (1992) 2863; 
Phys. Rev. B {\bf 48} (1993) 10345.
\bibitem{DMRG2} {\it Density-Matrix  Renormalization
- A new numerical method in physics -,} eds.
I.~Peschel, X.~Wang, M.~Kaulke and K.~Hallberg, (Springer Berlin, 1999),
and references there in.
\bibitem{DMRG3} U. Schollw\"ock: Rev. Mod. Phys. {\bf 77} (2005) 259.
\bibitem{Ostlund} S.~\"Ostlund and S.~Rommer: Phys. Rev. Lett {\bf 75} (1995) 3537.
\bibitem{Ostlund2} S.~Rommer and S.~\"Ostlund: Phys. Rev. B {\bf 55} (1997) 2164.
\bibitem{Ostlund3} M.~Andersson, M.~Boman, and S.~\"Ostlund: Phys. Rev. B {\bf 5}9 (1999) 10493.
\bibitem{Takasaki} H.~Takasaki, T.~Hikihara, and T.~Nishino: J. Phys. Soc. Jpn. {\bf 68} (1999) 1537.
\bibitem{Sierra}  J.~Dukelsky, M.A.~Mart\'in-Delgado, T.~Nishino and G.~Sierra: 
Europhys. Lett. {\bf 43} (1998) 457.
\bibitem{Acce} S.R.~White and I.~Affleck: Phys. Rev. B {\bf 54} (1996) 9862.
\bibitem{Acce2} S.R.~White: Phys Rev Lett. {\bf 77} (1996) 3633.
\bibitem{PWFRG} T.~Nishino and K.~Okunishi: J. Phys. Soc. Jpn. {\bf 64} (1995) 4084.
\bibitem{PWFRG2} K.~Ueda, T.~Nishino, K~Okunishi, Y.~Hieida, R.~Derian, and A.~Gendiar:
J. Phys. Soc. Jpn. {\bf 75} (2006) 014003.
\bibitem{Acts} N.~Akutsu and Y.~Akutsu: Phys. Rev. B {\bf 57} (1998) R4233;
N.~Akutsu and Y.~Akutsu: Prog. Theor. Phys. {\bf 105} (2001) 123.
\bibitem{Acts3} 
N.~Akutsu, Y.~Akutsu, and T.~Yamamoto: Prog. Theor. Phys. {\bf 105} (2001) 361;
N.~Akutsu, Y.~Akutsu, and T.~Yamamoto: 
Phys. Rev. B {\bf 64} (2001) 085415;
N.~Akutsu, Y.~Akutsu, and T.~Yamamoto: 
Journal of Crystal Growth {\bf 237-239} (2002) 14;
N.~Akutsu, Y.~Akutsu, and T.~Yamamoto:
Phys. Rev. B {\bf 67} (2003) 125407.
\bibitem{Hieida} Y.~Hieida, K.~Okunishi and Y.~Akutsu:
Phys. Lett. A {\bf 233} (1997) 464.
\bibitem{Hagiwara} 
M.~Hagiwara, Y.~Narumi, K.~Kindo, M.~Kohno, H.~Nakano, R.~Sato, and M.~Takahashi: 
Phys. Rev. Lett. {\bf 80} (1998) 1312.
\bibitem{Okunishi} K.~Okunishi, Y.~Hieida, and Y.~Akutsu: Phys. Rev. B {\bf 59}
(1999) 6806;
K.~Okunishi, Y.~Hieida, and Y.~Akutsu
Phys. Rev. E {\bf 59} (1999) R6227;
Y.~Hieida, K.~Okunishi, and Y.~Akutsu:
New Journal of Physics {\bf 1} (1999) 7.1;
K.~Okunishi, Y.~Hieida, and Y.~Akutsu: 
Phys. Rev. B {\bf 60} (1999) R6953;
Y.~Hieida, K.~Okunishi, and Y.~Akutsu: 
Phys. Rev. B {\bf 64} (2001) 224422.
\bibitem{Narumi} Y.~Narumi, K.~Kindo, M.~Hagiwara, H.~Nakano, A.~Kawaguchi
K.~Okunishi, and M.~Kohno: 
Phys. Rev. B {\bf 69} (2004) 174405.
\bibitem{Yoshikawa} S.~Yoshikawa, K.~Okunishi, M.~Senda and S.~Miyashita: 
J. Phys. Soc. Jpn. {\bf 73} (2004) 1798.
\bibitem{McClloch} I.~McClloch: arXiv: 0804.2509.
\bibitem{small} It is possible to choose the case $2N = 0$ or $2N = 2$ as the
starting point of DMRG calculation, where the choice is interesting from the
computational view point.
\bibitem{int} T.~Nishino, T.~Hikihara, K.~Okunishi, and Y.~Hieida: Int. J. Mod. 
Phys. B {\bf 13} (1999) 1.
\bibitem{commun} We are informed that the 2-site shift scheme is used in the
evaluation of the PWFRG method in Ref.[32], where dimension of the center matrix is 
restricted to $m$. 
\end{thebibliography}
\end{document}